\documentclass[11pt,draftcls,onecolumn]{IEEEtran}
\IEEEoverridecommandlockouts

\usepackage{amsmath}
\usepackage{amssymb}
\usepackage{graphicx}
\usepackage{psfrag}
\usepackage{url}
\usepackage{cite}
\usepackage{algorithm}
\usepackage{algpseudocode}
\usepackage{cases}

\newcommand{\E}{\mathrm{E}}
\newcommand{\VAR}{\mathrm{VAR}}

\DeclareMathOperator{\dB}{dB}
\DeclareMathOperator{\CSIT}{CSIT}
\DeclareMathOperator{\qCSIT}{q-CSIT}
\DeclareMathOperator{\Prob}{Pr}
\DeclareMathOperator{\out}{out}
\DeclareMathOperator{\rlz}{rlz}

\providecommand{\abs}[1]{\lvert{#1}\rvert}
\providecommand{\asq}[1]{\abs{#1}^2}

\begin{document}

\title{Distortion Minimization in Gaussian Layered Broadcast Coding with Successive Refinement}

\author{Chris~T.~K.~Ng,~\IEEEmembership{Member,~IEEE,}
        Deniz~G\"{u}nd\"{u}z,~\IEEEmembership{Member,~IEEE,}\\
        Andrea~Goldsmith,~\IEEEmembership{Fellow,~IEEE}
        and~Elza~Erkip,~\IEEEmembership{Senior~Member,~IEEE}
\thanks{This work was supported by the US Army under MURI award W911NF-05-1-0246,
the ONR under award N00014-05-1-0168, DARPA's ITMANET program under grant 1105741-1-TFIND,
a grant from Intel,
and the NSF under grant 0430885 and grant 0635177.
Chris~T.~K.~Ng was supported by a Croucher Foundation Fellowship.
The material in this paper was presented in part at the
IEEE International Conference on Communications, Glasgow, Scotland, UK, June 2007,
and at the IEEE International Symposium on Information Theory, Nice, France, June 2007.}%
\thanks{Chris~T.~K.~Ng is with the Bell Laboratories, Alcatel-Lucent, Holmdel, NJ 07733 USA (e-mail: Chris.Ng@alcatel-lucent.com).
Andrea~Goldsmith is with the Department of Electrical Engineering, Stanford University, Stanford, CA 94305 USA (e-mail: andrea@wsl.stanford.edu).
Deniz~G\"{u}nd\"{u}z is with the Department of Electrical Engineering, Stanford University, Stanford, CA 94305 USA,
and the Department of Electrical Engineering, Princeton University, Princeton, NJ 08544, USA (e-mail: dgunduz@princeton.edu).
Elza~Erkip is with the Department of Electrical and Computer Engineering, Polytechnic Institute of New York University, Brooklyn, NY 11201 USA (e-mail: elza@poly.edu).}%
}

\maketitle


\begin{abstract}

A transmitter without channel state information (CSI) wishes to send a delay-limited Gaussian source over a slowly fading channel.
The source is coded in superimposed layers, with each layer successively refining the description in the previous one.
The receiver decodes the layers that are supported by the channel realization and reconstructs the source up to a distortion.
The expected distortion is minimized by optimally allocating the transmit power among the source layers.
For two source layers, the allocation is optimal when power is first assigned to the higher layer up to a power ceiling that depends only on the channel fading distribution; all remaining power, if any, is allocated to the lower layer.
For convex distortion cost functions with convex constraints, the minimization is formulated as a convex optimization problem.
In the limit of a continuum of infinite layers, the minimum expected distortion is given by the solution to a set of linear differential equations in terms of the density of the fading distribution.
As the bandwidth ratio $b$ (channel uses per source symbol) tends to zero, the power distribution that minimizes expected distortion converges to the one that maximizes expected capacity.
While expected distortion can be improved by acquiring CSI at the transmitter (CSIT) or by increasing diversity from the realization of independent fading paths, at high SNR the performance benefit from diversity exceeds that from CSIT, especially when $b$ is large.

\end{abstract}


\IEEEpeerreviewmaketitle

\section{Introduction}
\label{sec:intro}


\PARstart{I}{n} an ergodic wireless channel, from the source-channel separation theorem \cite{shannon59:src_coding}, it is optimal to first compress the source and incur the associated distortion at a rate equal to the channel capacity, then send the compressed representation over the channel at capacity with asymptotically small error.
However, when delay constraints stipulate that the receiver decodes within a single realization of a slowly fading channel, without channel state information (CSI) at the transmitter, the transmission over a single fading block is non-ergodic and source-channel separation is not necessarily optimal.
In this case it is possible to reduce the end-to-end distortion of the reconstructed source by jointly optimizing the source-coding rate and the transmit power allocation based on the characteristics of the source and the channel.
In particular, we consider using the layered broadcast coding approach with successive refinement in the transmission of a Gaussian source over a slowly fading channel, in the absence of CSI at the transmitter.
First we assume the channel has a finite number of discrete fading states, then we extend the results to continuous fading distributions, for example, Rayleigh fading with diversity from the realization of independent fading paths.
The source is coded in layers, with each layer successively refining the description in the previous one.
The transmitter simultaneously transmits the codewords of all layers to the receiver by superimposing them with an appropriate power allocation.
The receiver successfully decodes the layers supported by the channel realization, and combines the descriptions in the decoded layers to reconstruct the source up to a distortion.
In this paper, we are interested in minimizing the expected distortion, and more generally, a convex distortion cost function, of the reconstructed source by optimally allocating the transmit power among the layers of codewords.
The system model is applicable to communication systems with real-time traffic where it is difficult for the transmitter to learn the channel condition.
For example, in a satellite voice system, it is desirable to consider the efficient transmission of the voice streams over uncertain channels that minimize the end-to-end distortion.

The broadcast strategy is proposed in \cite{cover72:broadcast_ch} to characterize the set of achievable rates when the channel state is unknown at the transmitter.
In the case of a Gaussian channel under Rayleigh fading,
\cite{shamai97:bc_strat_slow_fade} describes the layered broadcast coding approach, and derives the optimal power allocation that maximizes expected capacity when the channel has a single-antenna transmitter and receiver.
The layered broadcast approach is extended to multiple-antenna channels and the corresponding achievable rates are presented in \cite{shamai03:bc_app_slow_fade_mimo}.
In \cite{whiting06:bc_uncer_ch_dec_constr}, coding theorems are presented for the broadcast approach with delayed error-free feedback under decoding delay constraints.

In the transmission of a Gaussian source over a Gaussian channel, uncoded transmission is optimal \cite{goblick65:lim_tx_analog_src} in the special case when the source bandwidth equals the channel bandwidth \cite{gastpar03:to_code_or_not}.
For other bandwidth ratios, hybrid digital-analog joint source-channel transmission schemes are studied in \cite{shamai98:sys_lossy_src_ch, mittal02:hda_src_ch_bc_rc, reznic06:dstrn_bnd_bc_bw_exp}; in these works the codes are designed to be optimal at a target SNR but degrade gracefully should the realized SNR deviate from the target.
In particular, \cite{mittal02:hda_src_ch_bc_rc} conjectures that no code is simultaneously optimal at different SNRs when the source and channel bandwidths are not equal.
In this paper, the code considered is not targeted for a specific fading state; we minimize a convex distortion cost function over the fading distribution of the channel.

In \cite{laneman05:src_ch_parl_ch}, the minimum distortion is investigated in the transmission of a source over two independently fading channels in terms of the distortion exponent, which is defined as the exponential decay rate of the expected distortion in the high SNR regime.
Upper bounds on the distortion exponent and achievable joint source-channel schemes are presented in \cite{gunduz05:src_ch_code_quasi_fading} for a single-antenna quasi-static Rayleigh fading channel,
and later in \cite{gunduz08:jt_src_ch_code_mimo, caire05:snr_expn_hybrid_st} for multiple-antenna channels.
One of the proposed schemes in \cite{gunduz08:jt_src_ch_code_mimo}, layered source coding with progressive transmission (LS), is analyzed in terms of expected distortion for a finite number of layers at finite SNR in \cite{etemadi06:opt_layered_tx}.
The results in \cite{gunduz05:src_ch_code_quasi_fading, gunduz08:jt_src_ch_code_mimo} show that the broadcast strategy with layered source coding under an appropriate power allocation scheme is optimal for multiple-input single-output (MISO) and single-input multiple-output (SIMO) systems in terms of the distortion exponent.
Numerical optimization of the power allocation with constant rate among the layers is examined in \cite{sesia05:pro_sup_hyb}, while \cite{zachariadis05:src_fid_fading} considers
the optimization of power and rate allocation and presents approximate solutions in the high SNR regime.
Motivated by the optimality of the broadcast strategy in the high SNR regime, in this work, first presented in \cite{ng07:recur_pow_lbc, ng07:min_dist_lbc}, we investigate minimizing a convex distortion cost function, and in particular, the expected distortion, at any arbitrary finite SNR\@.

In a recent work in \cite{tian08:sr_bc_exp_dist_gaus}, the minimization of a linear distortion cost function, i.e., the expected distortion, is considered, and optimal power allocation algorithms are presented for discrete and continuous channel fading.
In this paper, we study how the properties of the optimal power allocation are affected by the channel-source bandwidth ratio, operating SNR, channel quantization, diversity order, and the related metric of capacity maximization.
Moreover, when the channel has discrete fading states, we also consider the minimization of an arbitrary convex distortion cost function under convex constraints, by formulating the distortion minimization as a convex optimization problem.
In minimizing the expected distortion, \cite{tian08:sr_bc_exp_dist_gaus} presents algorithms that calculate the optimal rate vector and power allocation based on a linear distortion cost function; general convex cost functions and constraints on the distortion realizations are not considered.
We show that the feasible distortion region is convex in layered broadcast coding with successive refinement, and the minimization of convex distortion cost functions can be efficiently solved with convex optimization numerical techniques.
The minimization of a general convex distortion cost function under continuous channel fading distributions, however, remains an open research problem.

The remainder of the paper is organized as follows.
The system model is presented in Section~\ref{sec:sys_mod}, and the layered broadcast coding scheme with successive refinement is explained in more detail in Section~\ref{sec:layered_bc_code}.
Section~\ref{sec:two_layers} focuses on the optimal power allocation between two layers, with the analysis being extended in Section~\ref{sec:recur_pow_alloc} to consider minimizing a convex distortion cost function over power allocation among multiple discrete layers.
The optimal power allocation for discretized Rayleigh fading distributions are presented in Section~\ref{sec:disc_ray_fading}.
Minimizing expected distortion under continuous fading distributions are treated in Section~\ref{sec:cont_fading} by studying the limiting process as the channel discretization resolution increases.
Section~\ref{sec:ray_div} considers the optimal power distribution and minimum expected distortion in Rayleigh fading channels with diversity, followed by conclusions in Section~\ref{sec:conclu}.

\section{System Model}
\label{sec:sys_mod}

\begin{figure}
  \centering
  \psfrag{r~f(r)}[][]{$\gamma \sim f(\gamma)$}
  \psfrag{xN}[][]{$x^N$}
  \psfrag{yN}[][]{$y^N$}
  \psfrag{sK}[][]{$s^K$}
  \psfrag{sKh}[][]{$\hat{s}^K$}
  \psfrag{CN(0,1)}[][]{$\mathcal{CN}(0,1)$}
  \includegraphics{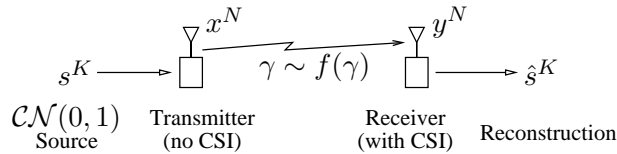}
  \caption{Source-channel coding without CSI at the transmitter.}
  \label{fig:src_ch_coding_pdf}
\end{figure}

Consider the system model illustrated in Fig.~\ref{fig:src_ch_coding_pdf}: A transmitter wishes to send a Gaussian source over a wireless channel to a receiver, at which the source is to be reconstructed up to a distortion.
Let the source be denoted by $s$, which is a sequence of independent identically distributed (iid) zero-mean circularly symmetric complex Gaussian (ZMCSCG) random variables with unit variance: $s\in\mathbb{C}\sim\mathcal{CN}(0,1)$.
The transmitter and the receiver each have a single antenna and the channel is described by
\begin{align}
  y &= Hx + n,
\end{align}
where $x\in\mathbb{C}$ is the transmit signal, $y\in\mathbb{C}$ is the received signal, and $n\in\mathbb{C}\sim\mathcal{CN}(0,1)$ is iid unit-variance ZMCSCG noise.

Suppose the distribution of the channel power gain is described by the probability density function (pdf) $f(\gamma)$, where $\gamma \triangleq \asq{h}$ and $h\in\mathbb{C}$ is a realization of $H$.
We first consider fading distributions with a finite number of discrete fading states; subsequently we generalize to continuous fading distributions.
The receiver has perfect CSI but the transmitter has only channel distribution information (CDI), i.e., the transmitter knows the pdf $f(\gamma)$ but not its instantaneous realization.
The channel is modeled by a quasi-static block fading process: $H$ is realized iid at the onset of each fading block and remains unchanged over the block duration.
We assume decoding at the receiver is \emph{delay-limited};
namely, delay constraints preclude coding across fading blocks but dictate that the receiver decodes at the end of each block.
Hence the transmission over a single fading block is non-ergodic.

Suppose each fading block spans $N$ channel uses, over which the transmitter describes $K$ of the source symbols. We define the \emph{bandwidth ratio} as $b\triangleq N/K$, which relates the number of channel uses per source symbol.
At the transmitter there is a power constraint on the transmit signal $\E\bigl[\asq{x}\bigr] \leq P$, where the expectation is taken over repeated channel uses over the duration of each fading block.
We assume $K$ is large enough to consider the source as ergodic, and $N$ is large enough to design codes that achieve the instantaneous channel capacity of a given fading state with negligible probability of error.
At the receiver, the channel output $y$ is used to reconstruct an estimate $\hat{s}$ of the source.
The distortion $D$ is measured by the mean squared error $\E\bigl[\asq{s-\hat{s}}\bigr]$ of the estimator, where the expectation is taken over the $K$-sequence of source symbols and the noise distribution.
The instantaneous distortion of the reconstruction depends on the fading realization of the channel; we are interested in minimizing the expected distortion $\E[D]$, where the expectation is over the fading distribution, and more generally, a convex distortion cost function with convex constraints in terms of the possible distortion realizations.

\section{Layered Broadcast Coding with Successive Refinement}
\label{sec:layered_bc_code}


To characterize the set of achievable rates when the channel state is unknown at the transmitter, a broadcast strategy is described in \cite{cover72:broadcast_ch}.
The transmitter designs its codebook by imagining it is communicating with an ensemble of virtual receivers.
Each virtual receiver corresponds to a fading state:
the realization of the fading state is taken as the channel gain of the virtual receiver.
The realized rate at the original receiver is given by the decodable rate of the realized virtual receiver.
A fading channel without transmitter CSI, therefore, can be modeled as a broadcast channel (BC).
In particular, the capacity region of the BC defines the maximal set of achievable rates among the virtual receivers, which, in terms of the original fading channel, is the maximal set of realized rates among the fading states.
In this work, we derive the optimal operating point in the BC capacity region that minimizes a convex distortion cost function, and in particular, the expected distortion $\E[D]$, of the reconstructed source.


For fading Gaussian channels, a layered broadcast coding approach is described in \cite{shamai97:bc_strat_slow_fade, shamai03:bc_app_slow_fade_mimo}.
In the layered broadcast approach, the virtual receivers are ordered according to their channel strengths:
for single-antenna channels, the channel strength of a virtual receiver is given by the channel power gain of its corresponding fading state.
We interpret each codeword intended for a virtual receiver as a \emph{layer} of code,
and the transmitter sends the superposition of all layers to the virtual receivers.
The capacity region of a single-antenna Gaussian BC is achievable by successive decoding \cite{cover91:eoit}, in which each virtual receiver decodes, in addition to its own layer, all the layers below it (the ones with weaker channel strengths).
Hence each layer represents the \emph{additional} information over its lower layer that becomes decodable by the original receiver should the layer be realized.

The layered broadcast approach fits particularly well with the successive refinability \cite{equitz91:sus_refn_info, rimoldi94:sus_refn_info_ach} of a Gaussian source.
Successive refinability states that if a source is first described at rate $R_1$, then subsequently refined at rate $R_2$, the overall distortion is the same as if the source were described at rate $R_1+R_2$ in the first place.
As the Gaussian source is successively refinable, naturally, each layer in the broadcast approach can be used to carry refinement information of a lower layer.
Concatenation of broadcast channel coding with successive refinement source coding is shown in \cite{gunduz05:src_ch_code_quasi_fading, gunduz08:jt_src_ch_code_mimo} to be optimal in terms of the distortion exponent for MISO/SIMO systems.

\begin{figure}
  \centering
  \psfrag{p1:r1}[][]{$p_1:\gamma_1$}
  \psfrag{p2:r2}[][]{$p_2:\gamma_2$}
  \psfrag{pM:rM}[][]{$p_M:\gamma_M$}
  \psfrag{(P1,R1)}[][]{$(P_1,R_1)$}
  \psfrag{(P2,R2)}[][]{$(P_2,R_2)$}
  \psfrag{(PM,RM)}[][]{$(P_M,R_M)$}
  \psfrag{sK}[][]{$s^K$}
  \psfrag{sKh}[][]{$\hat{s}^K$}
  \includegraphics{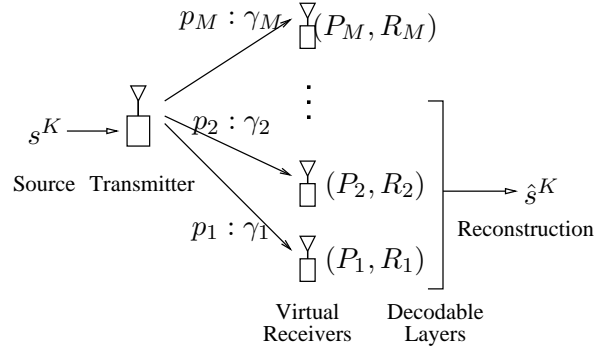}
  \caption{Layered broadcast coding with successive refinement.}
  \label{fig:src_ch_layers}
\end{figure}

We apply the layered broadcast approach and successive refinability to perform source-channel coding as outlined in Fig.~\ref{fig:src_ch_layers}.
First we assume the fading distribution has $M$ non-zero discrete states: the channel power gain realization is $\gamma_i>0$ with probability $p_i$, for $i=1,\dotsc,M$;
and we denote $p_0 \triangleq \Prob\{\gamma=0\}$.
Accordingly there are $M$ virtual receivers and the transmitter sends the sum of $M$ layers of codewords.
Let layer~$i$ denote the layer of codeword intended for virtual receiver~$i$,
and we order the layers as $\gamma_M>\dotsb>\gamma_1>0$.
We refer to layer~$M$ as the highest layer and layer~1 as the lowest layer.
Each layer successively refines the description of the source $s$ from the layer below it, and the codewords in different layers are independent.
Let $P_i$ be the transmit power allocated to layer~$i$, then the transmit symbol $x$ can be written as
\begin{align}
x &= \sqrt{P_1}\,x_1 + \sqrt{P_2}\,x_2 + \dotsb +\sqrt{P_M}\,x_M,
\end{align}
where $x_1,\dotsc,x_M$ are iid ZMCSCG random variables with unit variance.

With successive decoding, each virtual receiver first decodes and cancels the lower layers before decoding its own layer; the undecodable higher layers are treated as noise. Thus the rate $R_i$ (bits per channel use) intended for virtual receiver~$i$ is
\begin{align}
R_i &= \log\biggl(1+\frac{\gamma_i P_i}{1+\gamma_i \sum_{j=i+1}^{M}P_j}\biggr),\quad i=1,\dotsc,M,
\end{align}
where $\log$ is to base 2, and the term $\gamma_i \sum_{j=i+1}^{M}P_j$ represents the interference power from the higher layers.
Suppose $\gamma_k$ is the realized channel power gain, then the original receiver can decode layer~$k$ and all the layers below it. Hence the realized rate $R_{\rlz}^{(k)}$ at the original receiver is $R_1+\dotsb+R_k$.

From the rate distortion function of a complex Gaussian source \cite{cover91:eoit}, the mean squared distortion is $2^{-bR}$ when the source is described at a rate of $bR$ bits per symbol.
Thus the realized distortion $D_{\rlz}^{(k)}$ of the reconstructed source $\hat{s}$ is
\begin{align}
\label{eq:D_rlz_k_def}
  D_{\rlz}^{(k)} &= 2^{-b R_{\rlz}^{(k)}}
        = 2^{-b(R_1+\dotsb+R_k)},
\end{align}
where the last equality follows from successive refinability.
The expected distortion $\E[D]$ is obtained by averaging over the probability mass function (pmf) of the fading distribution:
\begin{align}
\label{eq:exp_dist_pR}
\E[D] &= \sum_{i=0}^{M} p_i D_{\rlz}^{(k)}
 = \sum_{i=0}^M p_i 2^{-b(\sum_{j=1}^i R_j)},
\end{align}
where $D_{\rlz}^{(0)} \triangleq 1$.

We begin by deriving the optimal power allocation $P_1^*,\dotsc,P_M^*$ among the layers to find the minimum expected distortion $\E[D]^*$.
Subsequently we generalize to consider minimizing a convex distortion cost function.
If a layer has an expected power gain of zero (i.e., $p_i\gamma_i=0$), the layer is allocated zero power;
hence in the derivation we assume $p_i\gamma_i\neq0$, for $i=1,\dotsc,M$.
Note that the expected distortion is monotonically decreasing in the transmit power $P$, hence the power constraint can be taken as an equality $\sum_{i=1}^{M} P_i = P$, and the optimization formulated as:
\begin{align}
\begin{split}
\E[D]^* &=
\min_{P_1,\dotsc,P_M}\E[D]\\
&\text{subject to } P_i\geq 0,\,{\textstyle\sum P_i}= P,\;\forall i=1,\dotsc,M.
\end{split}
\end{align}
We first consider the power allocation between two layers in the next section, then the analysis is extended to consider more than two layers in Section~\ref{sec:recur_pow_alloc}.
The layered source coding broadcast scheme can be straightforwardly extended to MISO/SIMO systems.
The equivalent single-antenna channel distribution is found by using isotropic inputs at the transmitter for MISO systems, and performing maximal-ratio combining at the receiver for SIMO systems.

\section{Two-Layer Optimal Power Allocation}
\label{sec:two_layers}

\begin{figure}
  \centering
  \psfrag{w:be}[][]{$w:\beta$}
  \psfrag{v:al}[][]{$u:\alpha$}
  \psfrag{(Ti-Ti1,Ri)}[][]{$(T_1-T_2,R_1)$}
  \psfrag{(Ti1,Ri1)}[][]{$(T_2,R_2)$}
  \psfrag{(Ti)}[][]{$(T_1)$}
  \includegraphics{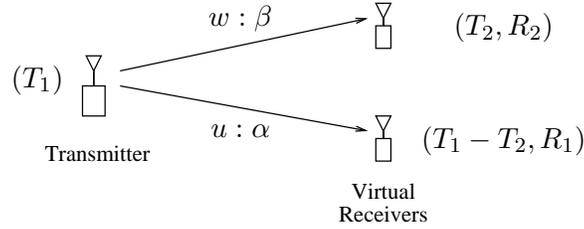}
  \caption{Power allocation between two layers.}
  \label{fig:two_var_layers}
\end{figure}

Suppose the channel fading distribution has only two states: the channel power gain realization is either $\alpha$ or $\beta$, with $\beta>\alpha>0$.
The transmitter then sends two layers ($M=2$) of codewords as shown in Fig.~\ref{fig:two_var_layers}.
Let $T_1$ denote the total transmit power constraint, and $T_2$ denote the power allocated to layer~2; the remaining power $T_1-T_2$ is allocated to layer~1.
The decodable rates for the virtual receivers are denoted by $R_1,R_2$; with successive decoding, they are given as follows:
\begin{align}
R_2 &= \log(1+\beta T_2)\\
R_1 &= \log\Bigl(1+\frac{\alpha(T_1-T_2)}{1+\alpha T_2}\Bigr).
\end{align}
Suppose we generalize slightly and consider the weighted distortion:
\begin{align}
\label{eq:two_wt_dist}
D_1 &= u 2^{-bR_1} + w 2^{-b(R_1+R_2)},
\end{align}
where the weights $\{u,w\}$ are non-negative. Note that the weighted distortion $D_1$ is the expected distortion $\E[D]$ when the weights $\{u,w\}$ are the probabilities of the fading realizations.

Given $T_1$, the total power available to the two layers, we optimize over $T_2$ to minimize the weighted distortion:
\begin{align}
D_1^* & = \min_{T_2\in[0,T_1]} D_1\\
 \label{eq:two_min_D_form}
 &= \min_{T_2\in[0,T_1]}
 \Bigl(\frac{1+\alpha T_1}{1+\alpha T_2}\Bigr)^{-b}
 \Bigl[u+(1+\beta T_2)^{-b}w\Bigr].
\end{align}
The minimization can be solved by the Lagrange method. We form the Lagrangian:
\begin{align}
L(T_2,\lambda_1,\lambda_2) & = D_1 + \lambda_1(T_2-T_1) - \lambda_2 T_2.
\end{align}
Applying the Karush-Kuhn-Tucker (KKT) necessary conditions, the gradient of the Lagrangian vanishes at the optimal power allocation $T_2^*$.
Specifically, the KKT conditions stipulate that at $T_2^*$, either one of the inequality constraints is active, or $d D_1/d T_2=0$.
Only one solution satisfies the KKT conditions, which leads to the optimal power allocation:
\begin{subnumcases}
  {\label{eq:T_i1_opt}T_2^* = \min(U_2,T_1) = }
  \label{eq:T_i1_opt_uncon}
  U_2 & if $U_2 \leq T_1$\IEEEeqnarraynumspace\\
  \label{eq:T_i1_opt_con}
  T_1 & else,
\end{subnumcases}
where
\begin{align}
\label{eq:U_i1}
U_2 &\triangleq
  \begin{cases}
    0 \hfill\text{if $\beta/\alpha \leq 1+u/w$}\\
    \dfrac{1}{\beta}\biggl(\Bigl[\dfrac{w}{u}
        \Bigl(\dfrac{\beta}{\alpha}-1\Bigr)\Bigr]^{\frac{1}{1+b}}-1\biggr) \qquad\qquad\text{else}.
  \end{cases}
\end{align}
In Section~\ref{sec:cvx_dist_cost_fn}, we show that the distortion minimization can be posed as a convex optimization problem; hence the KKT conditions are necessary and sufficient for optimality (we assume the given total power is non-zero so Slater's condition holds).
Interestingly, $U_2$ depends only on the layer parameters $w,\beta,u,\alpha$ (which are derived from the channel fading distribution) and the bandwidth ratio $b$, but not on the total power $T_1$.
In other words, the higher layer is allocated a \emph{fixed} amount of power as long as there is sufficient power available.
The optimal power allocation, therefore, adopts a simple policy: first assign power to the higher layer up to a ceiling of $U_2$, then assign all remaining power, if any, to the lower layer.

Under optimal power allocation $T_2^*$, the minimum weighted distortion as a function of the total power $T_1$ is given by
\begin{subnumcases}
  {\label{eq:min_wt_dist}D_i^* =}
  \label{eq:min_wt_dist_uncon}
  (1+\alpha T_1)^{-b}W_1 & if $U_2 \leq T_1$\\
  \label{eq:min_wt_dist_con}
  u + (1+\beta T_1)^{-b}w & else,
\end{subnumcases}
where
\begin{align}
W_1 &\triangleq (1+\alpha U_2)^b\bigl[u+(1+\beta U_2)^{-b}w\bigr].
\end{align}
Note that when the total power constraint $T_2\leq T_1$ is not active (\ref{eq:min_wt_dist_uncon}), the consequent minimum weighted distortion is analogous to that of a \emph{single} layer with channel power gain $\alpha$ and an equivalent weight $W_1$.
On the other hand, when the total power constraint is active (\ref{eq:min_wt_dist_con}),
it is equivalent to one with channel power gain $\beta$ and an equivalent weight $w$ (with an additive constant $u$ in the distortion).
Hence under optimal power allocation, with respect to the minimum expected distortion, the two layers can be represented by a single \emph{aggregate} layer; this idea is explored further when we consider multiple layers in Section~\ref{sec:recur_pow_alloc}.

\begin{figure}
  \centering
  \includegraphics*[width=8cm]{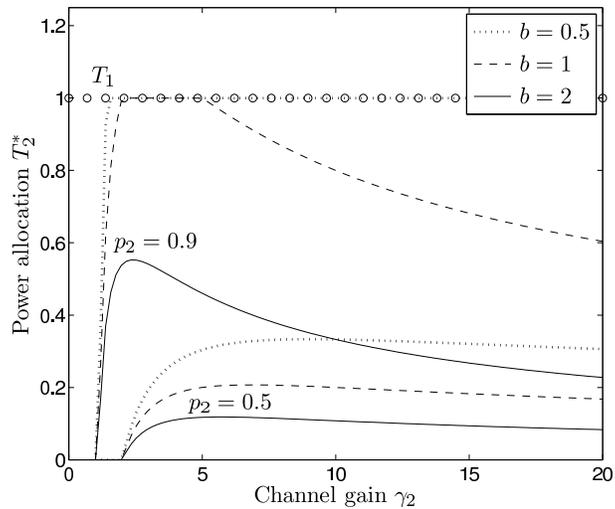}
  \caption{Optimal power allocation between two layers ($T_1=0~\dB$).}
  \label{fig:repa2_P2_r2}
\end{figure}

\begin{figure}
  \centering
  \includegraphics*[width=8cm]{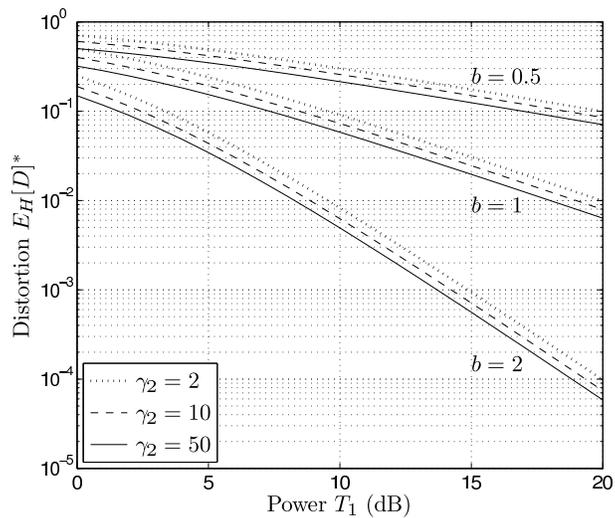}
  \caption{Two-layer minimum expected distortion ($p_2=0.5$).}
  \label{fig:repa2_ED_P}
\end{figure}

The optimal power allocation and the minimum expected distortion for a channel that has two discrete fading states is shown in Fig.~\ref{fig:repa2_P2_r2} and Fig.~\ref{fig:repa2_ED_P}, respectively, where  $T_2^*$, the power assigned to the top layer, is computed under the parameters:
\begin{align}
\begin{aligned}
  w &= p_2,\\
  u &= 1-p_2,
\end{aligned}
\qquad\qquad
\begin{aligned}
  \beta &= \gamma_2,\\
  \alpha &= 1.
\end{aligned}
\end{align}
Fig.~\ref{fig:repa2_P2_r2} shows that when the channel power gain $\gamma_2$ is small, an increase in $\gamma_2$ leads to a larger power allocation $T_2^*$ at the top layer, up to the total available power $T_1$. However, as $\gamma_2$ further increases, $T_2^*$ begins to fall.
This is because the transmission power in the higher layer is in effect interference to the lower layer.
When the top layer has a strong channel, the overall expected distortion is dominated by the bottom layer; in which case it is more beneficial to distribute the power to minimize the interference.
Fig.~\ref{fig:repa2_ED_P} plots the two-layer minimum expected distortion on a logarithmic scale, and it shows that the relative power gain of the channels has only a marginal impact on the expected distortion, as the overall distortion is in general dominated by the weaker channel.

\section{Multiple-Layer Power Allocation}
\label{sec:recur_pow_alloc}

In this section we consider the case when the fading distribution has $M$ fading states as depicted in Fig.~\ref{fig:src_ch_layers}, where $M$ is finite and $M\geq 2$.
For notational convenience, we write the power assignment as a cumulative sum starting from the top layer:
\begin{align}
\label{eq:T_j_def}
T_j \triangleq \sum_{i=j}^{M}P_i,\quad\text{for $j=1,\dotsc,M$}.
\end{align}
The original power assignments $\{P_1,\dotsc,P_M\}$ can then be recovered from $\{T_1,\dotsc,T_M\}$ by taking their differences.
By definition, $T_1 = P$ is given; hence the optimization is over the variables $T_2,\dots,T_M$:
\begin{align}
\begin{split}
\E[D]^* &=
\min_{T_2,\dotsc,T_M}\E[D]\\
&\text{subject to } 0\leq T_M\leq\dotsb\leq T_2\leq P.
\end{split}
\end{align}

\subsection{Expected Distortion Recurrence Relations}

In terms of the cumulative power variables $T_1,\dotsc,T_M$, the expected distortion in (\ref{eq:exp_dist_pR}) can be written as
\begin{align}
\label{eq:sum_cumprods}
\E[D] &= \sum_{i=1}^M p_i \Bigl(\prod_{j=1}^i \frac{1+\gamma_j T_j}{1+\gamma_j T_{j+1}}\Bigr)^{-b},
\end{align}
where $T_{M+1}\triangleq 0$.
We factor the sum of cumulative products in (\ref{eq:sum_cumprods}) and rewrite the expected distortion as a set of recurrence relations:
\begin{align}
D_M &\triangleq \bigl(1+\gamma_M T_M\bigr)^{-b}p_M\\
D_i &= \Bigl(\frac{1+\gamma_i T_i}{1+\gamma_i T_{i+1}}\Bigr)^{-b}\bigl(p_i+D_{i+1}\bigr),
\end{align}
where $i$ runs from $M-1$ down to 1.
We refer to $D_i$ as the cumulative distortion, which represents the cumulative effects on the expected distortion from layers~$i$ and above, with $D_1 = \E[D]$.
Note that given $D_{i+1}$ from the previous recurrence step, the term $D_i$ depends on only two adjacent power allocation variables $T_i$ and $T_{i+1}$; therefore, in each recurrence step $i$, we solve for the optimal $T_{i+1}^*$ in terms of $T_i$:
\begin{align}
D_M^* &\triangleq D_M\\
\label{eq:recur_min_dist}
D_i^*
&=\min_{T_{i+1}\in[0,T_i]}
\Bigl(\frac{1+\gamma_i T_i}{1+\gamma_i T_{i+1}}\Bigr)^{-b}\bigl(p_i+D_{i+1}^*\bigr).
\end{align}
In the last recurrence step ($i=1$), the minimum expected distortion $\E[D]^*$ is then given by $D_1^*$.

\subsection{Reduction through Optimal Power Allocation}
\label{sec:reduction_opt_pow}

We consider the layers from top to bottom.
In each recurrence step, the minimum distortion $D_i^*$ in (\ref{eq:recur_min_dist}) can be found by optimally allocating power between two adjacent layers as described in Section~\ref{sec:two_layers}.
In the first recurrence step ($i=M-1$), we consider the power allocation between the topmost two layers.
The minimal distortion $D_M^*$ is found by setting the parameters in (\ref{eq:two_min_D_form}) to be:
\begin{align}
\begin{aligned}
  w_{M-1} &= p_M,\\
  u_{M-1} &= p_{M-1},
\end{aligned}
\qquad\quad
\begin{aligned}
  \beta_{M-1} &= \gamma_M,\\
  \alpha_{M-1} &= \gamma_{M-1},
\end{aligned}
\end{align}
where the subscripts on the layer parameters $w,\beta,u,\alpha$ designate the recurrence step.
In general, in recurrence step $i$, the power allocation between layer~$i$ and layer~$i+1$ can be found by the optimization:
\begin{align}
\label{eq:min_dist_i_i1}
D_i^*
 &= \min_{T_{i+1}\in[0,T_i]}
 \Bigl(\frac{1+\alpha_i T_i}{1+\alpha_i T_{i+1}}\Bigr)^{-b}
 \Bigl[u_i+(1+\beta_i T_{i+1})^{-b}w_i\Bigr],
\end{align}
the solution of which is given in (\ref{eq:min_wt_dist}):
\begin{subnumcases}
  {\label{eq:min_wt_dist_i1}D_i^* =}
  \label{eq:min_wt_dist_i1_uncon}
  (1+\alpha_i T_i)^{-b}W_i & if $U_{i+1} \leq T_i$\IEEEeqnarraynumspace\\
  \label{eq:min_wt_dist_i1_con}
  u_i + (1+\beta_i T_i)^{-b}w_i & else.
\end{subnumcases}

There are two cases to the solution of $D_i^*$.
In the first case, the power allocation is not constrained by the available power $T_i$, and we substitute (\ref{eq:min_wt_dist_i1_uncon}) in the recurrence relation (\ref{eq:recur_min_dist}) to find the minimum distortion in the next recurrence step $i-1$:
\begin{align}
\label{eq:dist_i1_uncon}
D_{i-1}^* = \min_{T_i\in[0,T_{i-1}]}
\Bigl(\frac{1+\gamma_{i-1}T_{i-1}}{1+\gamma_{i-1}T_i}\Bigr)^{-b}
\Bigl[p_{i-1}+(1+\alpha_i T_i)^{-b}W_i
\Bigr].
\end{align}
The minimization in (\ref{eq:dist_i1_uncon}) has the same form as the one in (\ref{eq:min_dist_i_i1}), but with the following parameters:
\begin{align}
\label{eq:layer_param_uncon}
\begin{aligned}
  w_{i-1} &= W_i,\\
  u_{i-1} &= p_{i-1},
\end{aligned}
\qquad\qquad
\begin{aligned}
  \beta_{i-1} &= \alpha_i,\\
  \alpha_{i-1} &= \gamma_{i-1}.
\end{aligned}
\end{align}
Hence the minimization can be solved the same way as in the last recurrence step.
In the second case, the power allocation is constrained by the available power $T_i$, thus we instead substitute (\ref{eq:min_wt_dist_i1_con}) in (\ref{eq:recur_min_dist}) in the next recurrence step $i-1$:
\begin{align}
\label{eq:dist_i1_con}
D_{i-1}^* =
\min_{T_i\in[0,T_{i-1}]}
\Bigl(\frac{1+\gamma_{i-1}T_{i-1}}
{1+\gamma_{i-1}T_i}\Bigr)^{-b}
\Bigl[p_{i-1}+ u_i+(1+\beta T_i)^{-b}w_i\Bigr],
\end{align}
which again has the same form as in (\ref{eq:min_dist_i_i1}), with the following parameters:
\begin{align}
\label{eq:layer_param_con}
\begin{aligned}
  w_{i-1} &= w_i,\\
  u_{i-1} &= p_{i-1}+u_i,
\end{aligned}
\qquad\quad
\begin{aligned}
  \beta_{i-1} &= \beta_i,\\
  \alpha_{i-1} &= \gamma_{i-1}.
\end{aligned}
\end{align}
Therefore, in each recurrence step, the two-layer optimization procedure described in Section~\ref{sec:two_layers} can be used to find the minimum distortion and the optimal power allocation between the current layer and the aggregate higher layer.

\subsection{Feasibility of Unconstrained Minimizer}

When we proceed to the next recurrence step, however, it is necessary to determine which set of parameters in (\ref{eq:layer_param_uncon}), (\ref{eq:layer_param_con}) should be applied.
Note that in the optimization in (\ref{eq:min_dist_i_i1}), if the available power $T_i$ is unlimited (i.e., $T_i=\infty$), then the optimal power allocation is $T_{i+1}^* = U_{i+1}$ as given in (\ref{eq:T_i1_opt_uncon}); hence $U_{i+1}$ is the \emph{unconstrained} minimizer of $D_i$.
Consequently, we can first assume the minimization in (\ref{eq:min_dist_i_i1}) is unconstrained by $T_i$ and its solution is given by (\ref{eq:min_wt_dist_i1_uncon}).
If the unconstrained allocation $U_{i+1}$ is found to be feasible, then it is indeed the optimal allocation.
On the other hand, if $U_{i+1}$ is subsequently shown to be infeasible,
then we backtrack to the minimization in (\ref{eq:min_dist_i_i1}) and adopt the \emph{constrained} solution given by (\ref{eq:min_wt_dist_i1_con}).
In this case, $T_{i+1}^* = T_i$ as given in (\ref{eq:T_i1_opt_con}), which implies layer~$i$ is inactive since $P_{i}^*=T_i-T_{i+1}^*=0$.

We ascertain the feasibility of $U_{i+1}$ by verifying that it does not exceed the available power allocation $T_i$ from the lower layer~$i$, which in turn depends on the power allocation $T_{i-1}$ from the next lower layer~$i-1$ and so on.
The procedure can be accomplished by the recursive algorithm shown in Algorithm~\ref{alg:lbc_repa} in the Appendix.
We start by allocating power between the topmost two layers (line~\ref{line:start}).
In each recursion step, we compute the unconstrained allocation $U$ (line~\ref{line:uncon_min}).
If $U$ does not exceed the total power $P$, we first assume it is feasible, and proceed in the recursion to find the power allocation $T_i^*$ from the lower layer (line~\ref{line:alloc_uncon}).
If $U$ turns out to be infeasible, then we repeat the allocation step with the constrained minimization parameters (line~\ref{line:alloc_con}).
The recursion continues until the bottom layer is reached (line~\ref{line:bottom_layer}).
In the best case, if the unconstrained allocations for all layers are feasible, the algorithm has complexity $O(M)$. In the worst case, if all unconstrained allocations are infeasible, each recursion step performs two power allocations and the algorithm has complexity $O(2^M)$.

\subsection{Convex Distortion Cost Function}
\label{sec:cvx_dist_cost_fn}

In the previous sections, we consider minimizing a linear objective function of the possible distortion realizations $D_{\rlz}^{(k)}$'s; in particular, we consider the expected distortion $\E[D] = \sum_{k=1}^{M}p_k D_{\rlz}^{(k)}$.
Analytical solutions are presented that characterize the optimal power allocation that minimizes the expected distortion.
However, the expected distortion $\E[D]$ does not capture a user's sensitivity regarding the uncertainty in the range of possible outcomes of the distortion realizations $D_{\rlz}^{(k)}$'s.
In this section, we present a numerical optimization framework in which a wider class of objective functions is permissible.
Specifically, we consider the minimization of a distortion cost function $J\bigl(D_{\rlz}^{(1)},\dotsc,D_{\rlz}^{(M)}\bigr)$, where $J(\cdot)$ is convex in $D_{\rlz}^{(1)},\dotsc,D_{\rlz}^{(M)}$.
We show that the distortion minimization can be formulated as a convex problem; hence its solution can be computed efficiently by standard numerical methods in convex optimization \cite{boyd04:convex_opt}.

In terms of the cumulative power variables $T_j$'s defined in (\ref{eq:T_j_def}), the distortion realization $D_{\rlz}^{(k)}$ given in (\ref{eq:D_rlz_k_def}) can be written as:
\begin{align}
\label{eq:D_rlz_k_T_j}
D_{\rlz}^{(k)} &= \prod_{j=1}^{k} \Bigl(\frac{1+\gamma_j T_j}{1+\gamma_j T_{j+1}}\Bigr)^{-b},\quad k=1,\dotsc,M,
\end{align}
where $T_{M+1} \triangleq 0$ as previously defined.
Note that in (\ref{eq:D_rlz_k_T_j}), $D_{\rlz}^{(k)}$ can be written in terms of $D_{\rlz}^{(k-1)}$ as follows:
\begin{align}
\label{eq:Drlz_k_Drlz_k1_Tj}
D_{\rlz}^{(k)} &= D_{\rlz}^{(k-1)} \Bigl(\frac{1+\gamma_k T_k}{1+\gamma_k T_{k+1}}\Bigr)^{-b},\quad k=1,\dotsc,M,
\end{align}
where again recall $D_{\rlz}^{(0)} \triangleq 1$.
Next, we rearrange (\ref{eq:Drlz_k_Drlz_k1_Tj}) and write:
\begin{align}
\label{eq:T_j_T_j_1_D_rlz_k_1}
T_j = \Bigl(T_{j+1} + \frac{1}{\gamma_j}\Bigr)\biggl(\frac{D_{\rlz}^{(j)}}{D_{\rlz}^{(j-1)}}\biggr)^{-1/b} - \frac{1}{\gamma_j},
\quad j=1,\dotsc,M,
\end{align}
which we expand recursively initializing from $j=1$ to arrive at an expression for the total power $T_1 \triangleq \sum_{i=1}^M P_i$, which is given by:
\begin{align}
T_1 = -\frac{1}{\gamma_1} + \sum_{i=1}^{M} \Bigl(\frac{1}{\gamma_i}-\frac{1}{\gamma_{i+1}}\Bigr) \bigl(D_{\rlz}^{(i)}\bigr)^{-1/b},
\end{align}
where $\gamma_{M+1} \triangleq \infty$.
Under the power constraint $\sum_{i=1}^M P_i \leq P$, the distortion cost function minimization problem can be formulated as:
\begin{align}
\label{eq:min_J_D_rlz_1_M}
&\text{minimize}\quad J\bigl(D_{\rlz}^{(1)},\dotsc,D_{\rlz}^{(M)}\bigr)\\
&\text{over}\quad D_{\rlz}^{(1)},\dotsc,D_{\rlz}^{(M)}\notag\\
&\text{subject to}\notag\\
\label{eq:gamma_D_rlz_i_P}
& \quad -\frac{1}{\gamma_1} + \sum_{i=1}^{M} \Bigl(\frac{1}{\gamma_i}-\frac{1}{\gamma_{i+1}}\Bigr) \bigl(D_{\rlz}^{(i)}\bigr)^{-1/b}
\leq P\\
\label{eq:D_rlz_M_1_1}
& \quad 0 \leq D_{\rlz}^{(M)} \leq \dotsb \leq D_{\rlz}^{(1)} \leq 1,
\end{align}
where (\ref{eq:gamma_D_rlz_i_P}) corresponds to the power constraint from (\ref{eq:T_j_T_j_1_D_rlz_k_1}),
and (\ref{eq:D_rlz_M_1_1}) corresponds to the realized rates $R_{\rlz}^{(k)}$'s being nonnegative in (\ref{eq:D_rlz_k_def}).
Note that the constraints (\ref{eq:gamma_D_rlz_i_P}), (\ref{eq:D_rlz_M_1_1}) are convex:
in (\ref{eq:gamma_D_rlz_i_P}), $\bigl(D_{\rlz}^{(i)}\bigr)^{-1/b}$ is convex,
and $1/\gamma_i - 1/\gamma_{i+1} > 0$, which follows from the system model assumptions
$\gamma_M > \dotsb > \gamma_1 > 0$.
Therefore, the feasible distortion region $\{D_{\rlz}^{(1)},\dotsc,D_{\rlz}^{(M)}\}$, as characterized by (\ref{eq:gamma_D_rlz_i_P})--(\ref{eq:D_rlz_M_1_1}), is convex.
The objective function $J\bigl(D_{\rlz}^{(1)},\dotsc,D_{\rlz}^{(M)}\bigr)$ in (\ref{eq:min_J_D_rlz_1_M}) is convex by assumption.
It follows that minimizing $J\bigl(D_{\rlz}^{(1)},\dotsc,D_{\rlz}^{(M)}\bigr)$ over a convex region is a convex optimization problem: it can be efficiently solved by standard convex optimization numerical methods.
For instance, the optimization problem above can be solved using the \texttt{CVX} software package \cite{grant08:cvx_dcp_web, grant08:graph_nonsm_cvx}.

For example, to characterize the user's sensitivity to the uncertainty in the realized distortion, we may consider a risk-sensitive distortion cost function:
\begin{align}
\label{eq:J_phi_D_rlz_1M}
J_{\varphi}\bigl(D_{\rlz}^{(1)},\dotsc,D_{\rlz}^{(M)}\bigr) \triangleq \E[D] + \varphi\VAR[D],
\end{align}
where $\VAR[D]$ denotes the distortion variance:
\begin{align}
\VAR[D] &= \E\bigl[\bigl(D-\E[D]\bigr)^2\bigr]\\
&= \sum_{k=0}^{M} p_k \Bigl(D_{\rlz}^{(k)} - \sum_{i=0}^{M} p_i D_{\rlz}^{(i)}\Bigr)^2.
\end{align}
Note that $J_{\varphi}(\cdot)$ is a convex function of $D_{\rlz}^{(1)},\dotsc,D_{\rlz}^{(M)}$.
In (\ref{eq:J_phi_D_rlz_1M}), $\varphi\geq0$ is a given scalar constant, which represents the risk-aversion parameter \cite{boyd04:convex_opt}.
Accordingly, we may specify a suitable value of $\varphi$ to model the user's willingness to trade off an increase in the expected distortion $\E[D]$ in return for a reduction in the distortion variance $\VAR[D]$.
Under the convex optimization formulation in (\ref{eq:min_J_D_rlz_1_M})--(\ref{eq:D_rlz_M_1_1}), we may additionally consider convex constraints on $D_{\rlz}^{(1)},\dotsc,D_{\rlz}^{(M)}$.
For example, to guarantee worst-case performance under unfavorable fading states, we may consider the following maximum distortion or variance constraints:
\begin{align}
\E[D] &\leq D_{\max}\\
\VAR[D] &\leq V_{\max}\\
D_{\rlz}^{(k)} & \leq D_{\max}^{(k)}, \quad k=1,\dotsc,M,
\end{align}
where $D_{\max}$, $V_{\max}$, and $D_{\max}^{(k)}$'s are given constants.
In Section~\ref{sec:disc_ray_fading}, numerical examples are presented where we minimize the risk-sensitive distortion cost function $J_{\varphi}(\cdot)$ under discretized Rayleigh fading.

\section{Discretized Rayleigh Fading Distribution}
\label{sec:disc_ray_fading}

In this section, we present numerical results produced by the multiple-layer power allocation algorithms described in Section~\ref{sec:recur_pow_alloc}.
In the examples, we assume the channel pmf is taken from a discretized Rayleigh fading distribution.
Specifically, for a channel under Rayleigh fading with unit power, the channel power gain $\gamma$ is exponentially distributed with unit mean, and its probability density function (pdf) is given by
\begin{align}
  f(\gamma) = e^{-\gamma},\quad\text{for $\gamma\geq 0$}.
\end{align}
We truncate the pdf at $\gamma=\Gamma$, quantize $\gamma$ into $M$ evenly spaced levels:
\begin{align}
\gamma_i \triangleq i\,\Gamma/M,\quad \text{for $i=1,\dotsc,M$},
\end{align}
with $\gamma_0 \triangleq 0$,
and discretize the probability distribution of $\gamma$ to the closest lower level $\gamma_i$:
\begin{align}
p_i &\triangleq \int_{i\,\Gamma/M}^{(i+1)\Gamma/M} f(\gamma)\,d\gamma,
\quad\text{for $i=0,\dotsc,M-1$}\\
p_M &\triangleq \int_{\Gamma}^{\infty} f(\gamma)\,d\gamma.
\end{align}
While it is possible to consider the optimal discretization of a fading distribution that minimizes expected distortion \cite{sesia05:pro_sup_hyb, zachariadis05:src_fid_fading, etemadi07:rate_pow_layered}, in this paper we assume the channel pmf is given and do not consider such a step.

\begin{figure}
  \centering
  \includegraphics*[width=8cm]{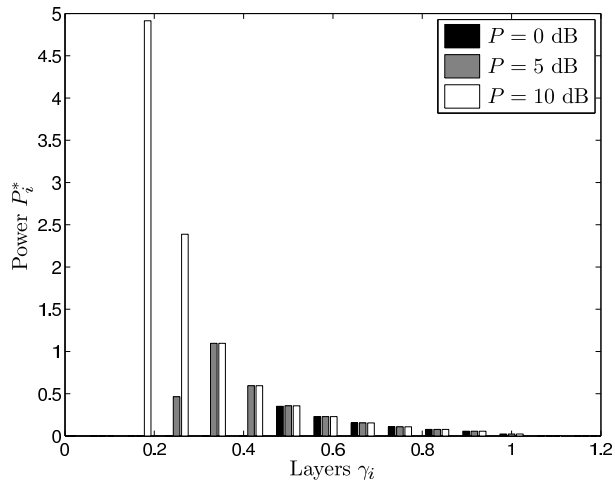}
  \caption{Optimal power allocation that minimizes the expected distortion $\E[D]$ ($b=1$).}
  \label{fig:Pi_P}
\end{figure}

\begin{figure}
  \centering
  \includegraphics*[width=8cm]{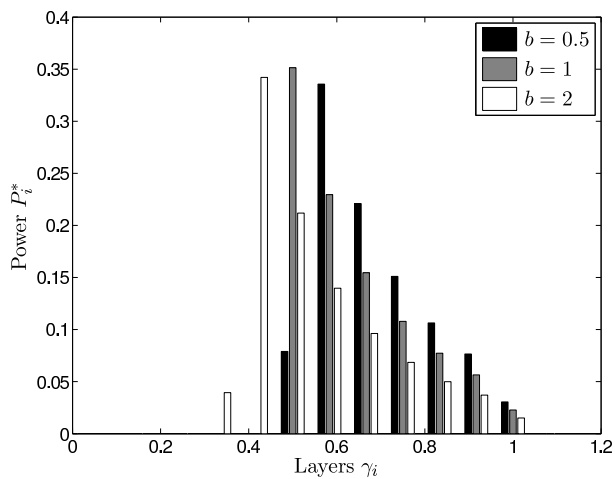}
  \caption{Optimal power allocation that minimizes the expected distortion $\E[D]$ ($P=0~\dB$).}
  \label{fig:Pi_b}
\end{figure}

The optimal power allocation that minimizes the expected distortion $\E[D]$ for the discretized Rayleigh fading pmf is shown in Fig.~\ref{fig:Pi_P} and Fig.~\ref{fig:Pi_b}. The Rayleigh fading pdf is truncated at $\Gamma=2$, and discretized into $M=24$ levels. The truncation is justified by the observation that in the output the highest layers near $\Gamma$ are not assigned any power.
Fig.~\ref{fig:Pi_P} plots the optimal power allocation $P_i^*$'s for different layers (indexed by the channel power gain $\gamma_i$) at SNRs $P=0~\dB$, $5~\dB$, and $10~\dB$, with the bandwidth ratio $b=1$.
We observe that the highest layers are inactive ($P_i^*=0$), and within the range of active layers a lower layer is in general allocated more power than a higher layer, except at the lowest active layer where it is assigned the remaining power.
As SNR increases, the power allocations of the higher layers are unaltered, but the range of active layers extends further into the lower layers.
On the other hand, Fig.~\ref{fig:Pi_b} plots the allocation $P_i^*$'s for different bandwidth ratios $b=0.5,1,2$ at the SNR of $0~\dB$.
It can be observed that a higher $b$ (i.e., more channel uses per source symbol) has the effect of spreading the power allocation further across into the lower layers.

Intuitively, the higher layers have stronger channels but suffer from larger risks of being in outage, while the lower layers provide higher reliability but at the expense of having to cope with less power-efficient channels.
Accordingly the optimal power allocation is concentrated around the middle layers.
Furthermore, as SNR increases, the numerical results suggest that, to minimize the expected distortion in a Rayleigh fading channel, it is more favorable to utilize the weaker channels with the extra power rather than accepting the larger outage risks from the higher layers.

\begin{figure}
  \centering
  \includegraphics*[width=8cm]{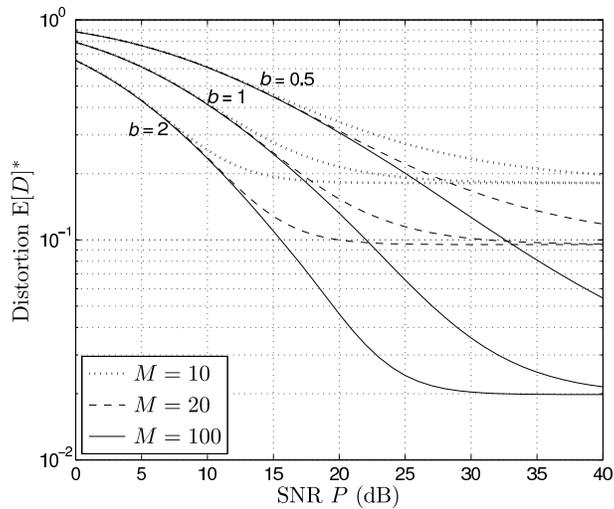}
  \caption{Minimum expected distortion under optimal power allocation.}
  \label{fig:ED_P_M_b}
\end{figure}

The minimum expected distortion $\E[D]^*$ under optimal power allocation is shown in Fig.~\ref{fig:ED_P_M_b} on a logarithmic scale.
When the bandwidth ratio $b$ is higher, $\E[D]^*$ decreases as expected.
However, the improvement in $\E[D]^*$ from refining the resolution $M$ of the discretization is almost negligible at low SNRs.
At high SNRs, on the other hand, the distortion is dominated by the outage probability:
\begin{align}
P_{\out} &\triangleq \Prob\{\gamma_0 = 0\text{ is realized}\}\\
&=\int_{0}^{\Gamma/M} f(\gamma)\,d\gamma,
\end{align}
which is decreasing in $M$.
Therefore, when the SNR is sufficiently high, the expected distortion $\E[D]^*$ reaches a floor that is dictated by $P_{\out}$.
This behavior is due to having evenly-spaced $\gamma_i$'s; the performance could be improved by optimizing the quantization level $\gamma_i$'s for the given $M$ layers.

\begin{figure}
  \centering
  \includegraphics*[width=8cm]{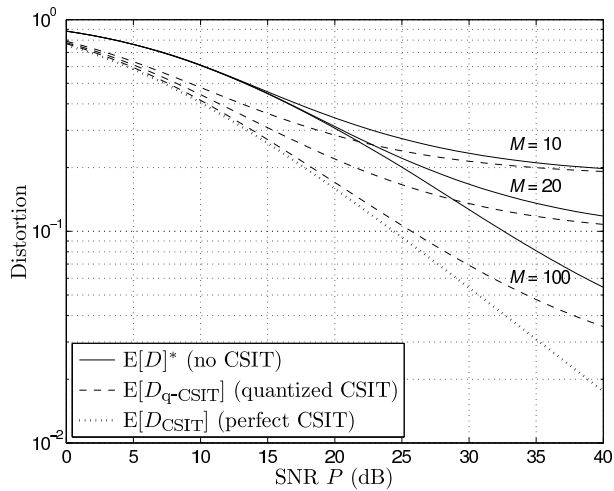}
  \caption{Expected distortion lower bounds with CSIT ($b=0.5$).}
  \label{fig:CSI_ray_dist}
\end{figure}

As a comparison, we consider the expected distortion lower bounds when the system has CSI at the transmitter (CSIT).
Under the discretized Rayleigh fading pmf, suppose the realized channel power gain is known to be $\gamma_k$,
then it is optimal for the transmitter to concentrate all power on layer~$k$ to achieve the instantaneous distortion $D_{\qCSIT} = (1+\gamma_k P)^{-b}$.
Thus with the quantized CSIT, the expected distortion is given by
\begin{align}
\E[D_{\qCSIT}] &= \sum_{k=0}^{M} p_k (1+\gamma_k P)^{-b}.
\end{align}
In terms of the original Rayleigh fading pdf $f(\gamma)$, with perfect CSIT, the expected distortion is similarly given by
\begin{align}
\E[D_{\CSIT}] &= \int_0^\infty e^{-\gamma}(1+\gamma P)^{-b}\,d\gamma,
\end{align}
where the definite integral can be evaluated numerically.
The expected distortions are plotted in Fig.~\ref{fig:CSI_ray_dist} for the cases of no CSIT, quantized and perfect CSIT\@.
It can be observed that at low SNRs, quantized CSIT is nearly as good as perfect CSIT,
whereas at high SNRs, quantized CSIT provides only marginal improvement over no CSIT, as the expected distortion is dominated by the probability of outage.

\begin{figure}
  \centering
  \includegraphics*[width=8cm]{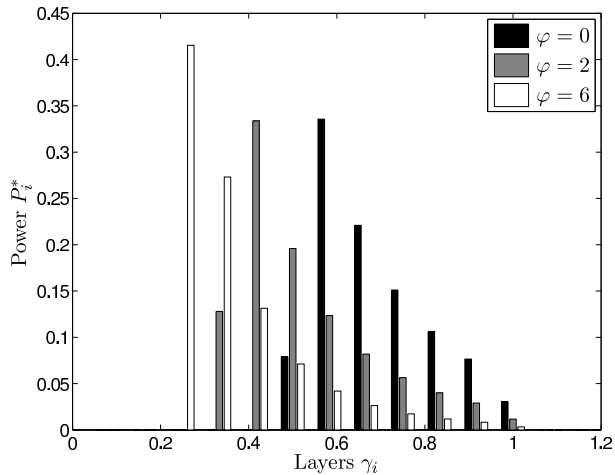}
  \caption{Optimal power allocation that minimizes $\E[D]+\varphi\VAR[D]$ ($P=0~\dB$, $b=0.5$).}
  \label{fig:J_p_EV}
\end{figure}

\begin{figure}
  \centering
  \includegraphics*[width=8cm]{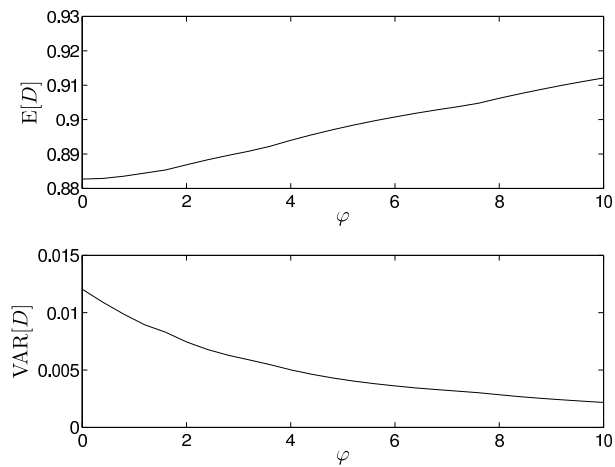}
  \caption{Expected distortion and variance corresponding to $\min \{\E[D]+\varphi\VAR[D]\}$ ($P=0~\dB$, $b=0.5$).}
  \label{fig:ED_VAR_rho}
\end{figure}

Numerical examples that illustrate the minimization of a convex cost function of the possible distortion realizations $D_{\rlz}^{(k)}$'s are shown in Fig.~\ref{fig:J_p_EV} and Fig.~\ref{fig:ED_VAR_rho}.
We consider the risk-sensitive distortion cost function: $J_{\varphi}\bigl(D_{\rlz}^{(1)},\dotsc,D_{\rlz}^{(M)}\bigr) \triangleq \E[D] + \varphi\VAR[D]$.
Fig.~\ref{fig:J_p_EV} shows the optimal power allocation that minimizes $J_{\varphi}(\cdot)$ for different values of the the risk-aversion parameter $\varphi$.
It is observed that a large $\varphi$, which represents the user's aversion to large variations in the realized distortion, shifts and concentrates the power allocation towards the lower layers.
Fig.~\ref{fig:ED_VAR_rho} plots the corresponding expected distortion $\E[D]$ and distortion variance $\VAR[D]$ that minimizes $J_{\varphi}(\cdot)$.
As $\varphi$ increases, it shows the tradeoff of accepting a higher $\E[D]$ for the reduction in $\VAR[D]$.

\section{Continuous Fading Distribution}
\label{sec:cont_fading}

In this section, we consider continuous fading distributions,
and we focus on the minimization of a linear distortion cost function, the expected distortion, by extending on the optimal power allocation analytical solutions derived in Section~\ref{sec:reduction_opt_pow}.
We study the limiting process as the discretization resolution of the fading distribution tends to infinity, and consider the optimal power distribution that minimizes the expected distortion when the fading distribution of the channel is given by a continuous probability density function.
Specifically, we assume the layers are evenly spaced, with $\gamma_{i+1}-\gamma_i = \Delta\gamma$, and we consider the limiting process as $\Delta\gamma\rightarrow0$ to obtain the power distribution:
\begin{align}
\rho(\gamma) \triangleq \lim_{\Delta\gamma \rightarrow 0}
\dfrac{1}{\Delta\gamma} P_{\lceil \gamma/\Delta\gamma \rceil},
\end{align}
where for discrete layers the power allocation $P_i$ is referenced by the integer layer index $i$, while the continuous power distribution $\rho(\gamma)$ is indexed by the channel power gain $\gamma$.

Consider the optimal power allocation between layer $\gamma$ and its next lower layer $\gamma-\Delta\gamma$.
Let $T(\gamma-\Delta\gamma)$ denote the available transmit power for layers~$\gamma-\Delta\gamma$ and above, of which $T(\gamma)$ is allocated to layers~$\gamma$ and above; the remaining power $T(\gamma-\Delta\gamma)-T(\gamma)$ is allocated to layer~$\gamma-\Delta\gamma$.
The optimal power allocation $T^*(\gamma)$ is given by the solution to the two-layer optimization problem in Section~\ref{sec:two_layers}, with the parameters in Fig.~\ref{fig:two_var_layers} correspondingly set to be:
\begin{align}
\begin{aligned}
  w &= W(\gamma),\\
  u &= f(\gamma)\Delta\gamma,
\end{aligned}
\qquad\qquad
\begin{aligned}
  \beta &= \gamma,\\
  \alpha &= \gamma-\Delta\gamma,
\end{aligned}
\end{align}
where $f(\gamma)$ is the pdf of the channel power gain with $f(\gamma)\Delta\gamma$ representing the probability that layer~$\gamma-\Delta\gamma$ is realized,
and $W(\gamma)$ is interpreted as an equivalent probability weight summarizing the aggregate effect of the layers~$\gamma$ and above.

From (\ref{eq:T_i1_opt}), (\ref{eq:U_i1}), the optimal power allocation is given by
\begin{subnumcases}
  {\label{eq:Tr_opt} T^*(\gamma) = }
  \label{eq:Tr_opt_uncon}
  U(\gamma) & if $U(\gamma) \leq T(\gamma-\Delta\gamma)$\IEEEeqnarraynumspace\\
  \label{eq:Tr_opt_con}
  T(\gamma-\Delta\gamma) & else,
\end{subnumcases}
where
\begin{subnumcases}
{\hspace{1em}U(\gamma) \triangleq}
    \label{eq:Ur_o}
    0 \text{\hspace{5.25em}if $\gamma \geq W(\gamma)/f(\gamma) + \Delta\gamma$} & \\
    \label{eq:Ur_nz}
    \dfrac{1}{\gamma}\biggl(\Bigl[\frac{W(\gamma)}{f(\gamma)(\gamma-\Delta\gamma)}\Bigr]^{\frac{1}{1+b}}-1\biggr)  \text{\hspace{2em}else.}&\hspace{1em}
\end{subnumcases}
We assume there is a region of $\gamma$ where the cumulative power allocation is not constrained by the power available from the lower layers, i.e., $U(\gamma)\leq U(\gamma-\Delta\gamma)$ and $U(\gamma)\leq P$.
In this region the optimal power allocation $T^*(\gamma)$ is given by the unconstrained minimizer $U(\gamma)$ in (\ref{eq:Tr_opt_uncon}).
In the solution to $U(\gamma)$ we need to verify that $U(\gamma)$ is non-increasing in this region, which corresponds to the power distribution $\rho^*(\gamma)$ being non-negative.
Following (\ref{eq:min_wt_dist_uncon}),
we write the cumulative distortion from layers $\gamma$ and above in the form:
\begin{align}
\label{eq:Dr_W_form}
D^*(\gamma) = \bigl(1+\gamma T(\gamma)\bigr)^{-b} W(\gamma).
\end{align}
Substitute in the unconstrained cumulative power allocation $U(\gamma)$,
the cumulative distortion at layer $\gamma-\Delta\gamma$ becomes:
\begin{align}
\label{eq:Drd_U_W}
D^*(\gamma-\Delta\gamma) &=
\Bigl(\frac{1+(\gamma-\Delta\gamma) T(\gamma-\Delta\gamma)}{1+(\gamma-\Delta\gamma) U(\gamma)}\Bigr)^{-b}
\Bigl[f(\gamma)\Delta\gamma+\bigl(1+\gamma U(\gamma)\bigr)^{-b} W(\gamma)\Bigr],
\end{align}
which is of the form in (\ref{eq:Dr_W_form}) if we define $W(\gamma-\Delta\gamma)$ by the recurrence equation:
\begin{align}
\label{eq:Wrd_Wr}
W(\gamma-\Delta\gamma) &=
\bigl(1+(\gamma-\Delta\gamma) U(\gamma)\bigr)^b
\bigl[f(\gamma)\Delta\gamma+\bigl(1+\gamma U(\gamma)\bigr)^{-b} W(\gamma)\bigr].
\end{align}

As the spacing between the layers condenses, in the limit of $\Delta\gamma$ approaching zero, the recurrence equations (\ref{eq:Drd_U_W}), (\ref{eq:Wrd_Wr}) become differential equations.
The optimal power distribution $\rho^*(\gamma)$ is given by the derivative of the cumulative power allocation:
\begin{align}
\rho^*(\gamma) &= -{T^*}'(\gamma),
\end{align}
where $T^*(\gamma)$ is described by solutions in three regions:
\begin{subnumcases}
{\label{eq:Tr_all_opt}T^*(\gamma) =}
\label{eq:Tr_ro}
    0 & $\gamma > \gamma_o$ \\
\label{eq:Tr_rP_ro}
    U(\gamma) & $\gamma_P \leq \gamma \leq \gamma_o$\\
\label{eq:Tr_rP}
    P & $\gamma < \gamma_P$.
\end{subnumcases}
In region (\ref{eq:Tr_ro}) when $\gamma > \gamma_o$, corresponding to cases (\ref{eq:Tr_opt_uncon}) and (\ref{eq:Ur_o}), no power is allocated to the layers and (\ref{eq:Wrd_Wr}) simplifies to $W(\gamma) = 1-F(\gamma)$, where $F(\gamma) \triangleq \int_0^\gamma f(s)\,ds$ is the cumulative distribution function (cdf) of the channel power gain.
The boundary $\gamma_0$ is defined by the condition in (\ref{eq:Ur_o}) which satisfies:
\begin{align}
\label{eq:ro_rf_F}
\gamma_o f(\gamma_o) + F(\gamma_o) - 1 = 0.
\end{align}
Under Rayleigh fading when $f(\gamma) = \bar{\gamma}^{-1}e^{-\gamma/\bar{\gamma}}$, where $\bar{\gamma}$ is the expected channel power gain, (\ref{eq:ro_rf_F}) evaluates to $\gamma_o = \bar{\gamma}$.
For other fading distributions, $\gamma_o$ may be computed numerically.

In region (\ref{eq:Tr_rP_ro}) when $\gamma_P \leq \gamma \leq \gamma_o$, corresponding to cases (\ref{eq:Tr_opt_uncon}) and (\ref{eq:Ur_nz}), the optimal power distribution is described by a set of differential equations.
We apply the first order binomial expansion $(1+\Delta\gamma)^b\cong1+b\Delta\gamma$, and (\ref{eq:Wrd_Wr}) becomes:
\begin{align}
W'(\gamma) &= \lim_{\Delta\gamma \rightarrow 0} \frac{W(\gamma) - W(\gamma-\Delta\gamma)}{\Delta\gamma}\\
\label{eq:dW_W}
&= b\frac{W(\gamma)}{\gamma} - (1+b)\Bigl[f(\gamma)\Big(\frac{W(\gamma)}{\gamma}\Bigr)^b\Bigr]^{\frac{1}{1+b}},
\end{align}
which we substitute in (\ref{eq:Ur_nz}) to obtain:
\begin{align}
\label{eq:dU_U}
U'(\gamma) &= -\Big(\frac{2/\gamma + f'(\gamma)/f(\gamma)}{1+b}\Bigr) \Big[U(\gamma)+1/\gamma\Bigr].
\end{align}
Hence $U(\gamma)$ is described by a first order linear differential equation.
With the initial condition $U(\gamma_o) = 0$, its solution is given by
\begin{align}
\label{eq:Ur_int_fr}
U(\gamma) &= \frac{\displaystyle -\int_{\gamma_o}^{\gamma}
\dfrac{1}{s}\Bigl(\dfrac{2}{s}+\dfrac{f'(s)}{f(s)}\Bigr) \bigl[s^2 f(s)\bigr]^{\frac{1}{1+b}} \,ds}
{(1+b)\bigl[\gamma^2 f(\gamma)\bigr]^{\frac{1}{1+b}}},
\end{align}
and condition (\ref{eq:Tr_opt_con}) in the lowest active layer becomes the boundary condition $U(\gamma_P) = P$.
In \cite{tian08:sr_bc_exp_dist_gaus}, the power distribution in (\ref{eq:Ur_int_fr}) is derived using the calculus of variations method.

Similarly, as $\Delta\gamma\rightarrow0$, the evolution of the expected distortion in (\ref{eq:Drd_U_W}) becomes:
\begin{align}
D'(\gamma) &= -\dfrac{b\gamma U'(\gamma)}{1+\gamma U(\gamma)}D(\gamma) - f(\gamma)\\
 &= \Bigl[\dfrac{b}{1+b}\Bigl(\dfrac{2}{\gamma}+\dfrac{f'(\gamma)}{f(\gamma)}\Bigr)\Bigr]D(\gamma)- f(\gamma),
\end{align}
which is again a first order linear differential equation.
With the initial condition $D(\gamma_o) = W(\gamma_o) = \gamma_o f(\gamma_o)$, its solution is given by
\begin{align}
D(\gamma) &= \frac{\displaystyle -\int_{\gamma_o}^{\gamma} f(s)
\Bigl[\Bigl(\dfrac{s}{\gamma_o}\Bigr)^2\dfrac{f(s)}{f(\gamma_o)}\Bigr]^{\frac{-b}{1+b}} \,ds
+ \gamma_o f(\gamma_o)}
{\Bigl[\Bigl(\dfrac{\gamma}{\gamma_o}\Bigr)^2\dfrac{f(\gamma)}{f(\gamma_o)}\Bigr]^{\frac{-b}{1+b}}}.
\end{align}
Under Rayleigh fading, for instance, the solutions to $U(\gamma)$ and $D(\gamma)$ are given by
\begin{align}
\label{eq:U_ray}
U(\gamma) &= \frac{\displaystyle \int_{\bar{\gamma}}^{\gamma}
\Bigl(\dfrac{1}{\bar{\gamma}}-\dfrac{2}{s}\Bigr) \bigl[s^{1-b}e^{-s/\bar{\gamma}}\bigr]^{\frac{1}{1+b}} \,ds
}
{(1+b)\bigl[\gamma^2 e^{-\gamma/\bar{\gamma}}\bigr]^{\frac{1}{1+b}}},
\end{align}
\begin{align}
\label{eq:D_ray}
D(\gamma) &= \frac{\displaystyle -\dfrac{1}{\bar{\gamma}}\int_{\bar{\gamma}}^{\gamma} e^{-s/\bar{\gamma}}
\Bigl[\Bigl(\dfrac{s}{\bar{\gamma}}\Bigr)^2 e^{-(s-\bar{\gamma})/\bar{\gamma}}\Bigr]^{\frac{-b}{1+b}} \,ds
+ e^{-1}}
{\Bigl[\Bigl(\dfrac{\gamma}{\bar{\gamma}}\Bigr)^2 e^{-(\gamma-\bar{\gamma})/\bar{\gamma}}\Bigr]^{\frac{-b}{1+b}}}.
\end{align}
The integrals in (\ref{eq:U_ray}), (\ref{eq:D_ray}) can be computed numerically by evaluating the incomplete gamma function.

Finally, in region (\ref{eq:Tr_rP}) when $\gamma < \gamma_P$, corresponding to case (\ref{eq:Tr_opt_con}), the transmit power $P$ has been exhausted, and no power is allocated to the remaining layers.
Hence the minimum expected distortion is
\begin{align}
\E[D]^* = D(0) = F(\gamma_P) + D(\gamma_P),
\end{align}
where the last equality follows from when $\gamma < \gamma_P$ in region (\ref{eq:Tr_rP}), $\rho^*(\gamma)=0$ and
$D(\gamma) = \int_\gamma^{\gamma_P} f(s) \,ds + D(\gamma_P)$.

\section{Rayleigh Fading with Diversity}
\label{sec:ray_div}

In this section we consider the optimal power distribution and the minimum expected distortion when the wireless channel undergoes Rayleigh fading with a diversity order of $L$ from the realization of independent fading paths.
Specifically, we assume the fading channel is characterized by the Erlang distribution:
\begin{align}
f_L(\gamma) = \frac{(L/\bar{\gamma})^L \gamma^{L-1} e^{-L\gamma/\bar{\gamma}} }
{(L-1)!}, \qquad\gamma \geq 0,
\end{align}
which corresponds to the average of $L$ iid channel power gains, each under Rayleigh fading with an expected value of $\bar{\gamma}$.
The $L$-diversity system may be realized by having $L$ transmit antennas using isotropic inputs, by relaxing the decode delay constraint over $L$ fading blocks, or by having $L$ receive antennas under maximal-ratio combining when the power gain of each antenna is normalized by $1/L$.

\begin{figure}
  \centering
  \includegraphics*[width=8cm]{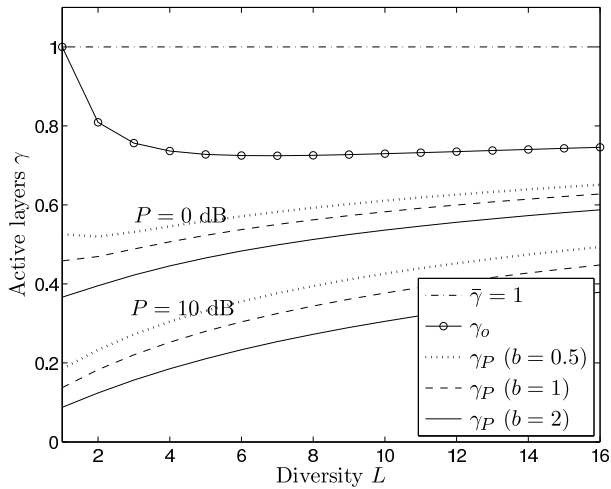}
  \caption{Span of active layers under optimal power distribution.}
  \label{fig:erlang_r0_rP}
\end{figure}

The optimal power distribution (\ref{eq:Tr_all_opt}) concentrates the transmit power over a range of active layers; the upper and lower boundaries $\gamma_o, \gamma_P$ of the span of the active layers are plotted in Fig.~\ref{fig:erlang_r0_rP}.
A higher SNR $P$ or a larger bandwidth ratio $b$ extends the span further into the lower layers but the upper boundary $\gamma_o$ remains unperturbed.
As $L$ increases, the fluctuation in the channel realization is diminished by the diversity of combining multiple independent fading paths, and the power distribution becomes more concentrated, albeit slowly.
By the law of large numbers, at asymptotically large $L$, we expect all power concentrates at $\bar{\gamma}$.

\begin{figure}
  \centering
  \includegraphics*[width=8cm]{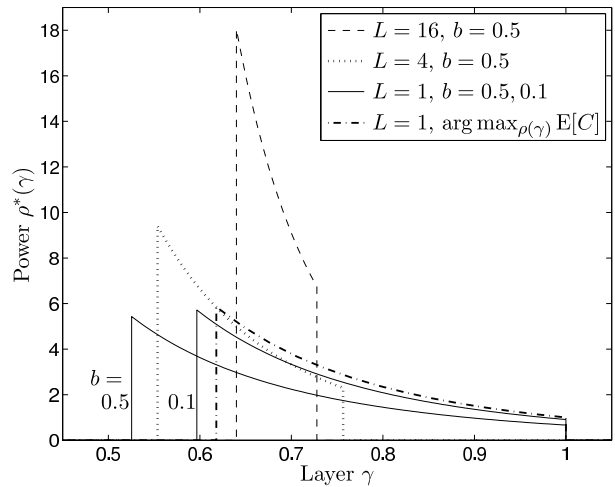}
  \caption{Optimal power distribution ($P=0~\dB$).}
  \label{fig:erlang_pow_dist}
\end{figure}

Fig.~\ref{fig:erlang_pow_dist} shows the optimal power allocation $\rho^*(\gamma)$.
It can be observed that a smaller bandwidth ratio $b$ reduces the spread of the power distribution.
In fact, as $b$ approaches zero, the optimal power distribution that minimizes expected distortion converges to the power distribution that maximizes expected capacity.
To show the connection, we take the limit in the distortion-minimizing cumulative power distribution in (\ref{eq:Ur_int_fr}):
\begin{align}
\lim_{b\rightarrow0}U(\gamma) &= \frac{1-F(\gamma)-\gamma f(\gamma)}{\gamma^2 f(\gamma)},
\end{align}
which is equal to the capacity-maximizing cumulative power distribution as derived in \cite{shamai03:bc_app_slow_fade_mimo}.
Essentially, from the first order expansion $e^b\cong1+b$ for small $b$, $\E[D]\cong 1- b \E[C]$ when the bandwidth ratio is small, where $\E[C]$ is the expected capacity in nats/s, and hence minimizing expected distortion becomes equivalent to maximizing expected capacity.
For comparison, the capacity-maximizing power distribution is also plotted in Fig.~\ref{fig:erlang_pow_dist}.
Note that the distortion-minimizing power distribution is more conservative, and it is more so as $b$ increases, as the allocation favors lower layers in contrast to the capacity-maximizing power distribution.

\begin{figure}
  \centering
  \includegraphics*[width=8cm]{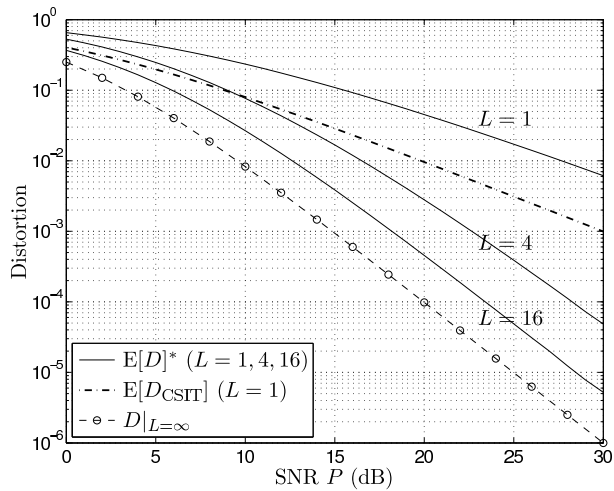}
  \caption{Minimum expected distortion ($b=2$).}
  \label{fig:erlang_ED_P}
\end{figure}

Fig.~\ref{fig:erlang_ED_P} shows the minimum expected distortion $\E[D]^*$ versus SNR for different diversity orders.
With infinite diversity, the channel power gain becomes constant at $\bar{\gamma}$, and the distortion is given by
\begin{align}
D\vert_{L=\infty} = (1+\bar{\gamma}P)^{-b}.
\end{align}
In the case when there is no diversity ($L=1$), a lower bound to the expected distortion is also plotted.
The lower bound assumes the system has CSI at the transmitter (CSIT),
which allows the transmitter to concentrate all power at the realized layer to achieve the expected distortion:
\begin{align}
\E[D_{\CSIT}] &= \int_0^\infty e^{-\gamma}(1+\gamma P)^{-b}\,d\gamma.
\end{align}
Note that at high SNR, the performance benefit from diversity exceeds that from CSIT, especially when the bandwidth ratio $b$ is large.
In particular, in terms of the distortion exponent $\Delta$ \cite{laneman05:src_ch_parl_ch},
it is shown in \cite{gunduz08:jt_src_ch_code_mimo} that in a MISO or SIMO channel, layered broadcast coding achieves:
\begin{align}
\Delta \triangleq - \lim_{P\rightarrow\infty} \frac{\log \E[D]}{\log P}
       = \min(b,L),
\end{align}
where $L$ is the total diversity order from independent fading blocks and antennas.
Moreover, the layered broadcast coding distortion exponent is shown to be optimal and CSIT does not improve $\Delta$, whereas diversity increases $\Delta$ up to a maximum as limited by the bandwidth ratio $b$.

\section{Conclusions}
\label{sec:conclu}

We considered the problem of source-channel coding over a delay-limited fading channel without CSI at the transmitter, and derived the optimal power allocation that minimizes the end-to-end distortion in the layered broadcast coding transmission scheme with successive refinement.
In the allocation of transmission power between two layers of codewords in a two-state fading channel, the optimal allocation that minimizes the expected distortion has a particular structure that lends itself to be generalized to the cases when the channel has multiple discrete fading states or a continuous fading distribution.
Specifically, the optimal two-layer allocation assigns power first to the higher layer, up to a power ceiling that depends only on the channel fading distribution but independent of the total available power; any surplus over the power ceiling is allocated to the lower layer.
When the channel has multiple discrete fading states, we write the minimum expected distortion as a set of recurrence relations, and in each recurrence step the two-layer optimization procedure solves the power allocation between the current layer and the aggregate higher layer.
The optimization framework is extended to consider convex distortion cost functions with convex constraints by posing the minimization as a convex optimization problem.
We applied the power allocation algorithms to the pmf of a discretized Rayleigh fading distribution.
We observed that the optimal power allocation is concentrated around the middle layers,
and within this range the lower layers are assigned more power than the higher ones.
As the SNR increases, the allocations of the higher layers remain unchanged, and the extra power is allocated to the idle lower layers.
The distortion-minimizing power distribution, therefore, is conservative:
it is more beneficial to utilize the lower layers, despite their weaker channel gains, than the higher layers as the latter have larger risks of being in outage.

We also derived the optimal power distribution that minimizes the expected distortion when the fading distribution of the channel is given by a continuous probability density function.
We computed the optimal power distribution for Rayleigh fading channels with diversity order $L$, and showed that increasing the diversity $L$ concentrates the power distribution towards the expected channel power gain $\bar{\gamma}$, while a larger bandwidth ratio $b$ spreads the power distribution further into the lower layers.
On the other hand, in the limit as $b$ tends to zero, the optimal power distribution that minimizes expected distortion converges to the power distribution that maximizes expected capacity.
While the expected distortion can be improved by acquiring CSIT or increasing the diversity order, it is shown that at high SNR the performance benefit from diversity exceeds that from CSIT, especially when the bandwidth ratio $b$ is large.
Under continuous channel fading, in this paper we focused on minimizing the expected distortion, which is a linear cost function of the distortion realizations.
Future research works may include considering the minimization of a general convex distortion cost function under continuous channel fading distributions.

\appendix

\begin{algorithm}[H]
\caption{Multiple-Layer Power Allocation}
\label{alg:lbc_repa}
\begin{algorithmic}[1]

\State \Call{alloc}{$M-1,p_M,\gamma_M,p_{M-1},\gamma_{M-1}$} \Comment{Start from top}
  \label{line:start}

\Statex
\Procedure{alloc}{$i,w,\beta,u,\alpha$}
  \State Compute $U$ from $w,\beta,u,\alpha$
    \label{line:uncon_min}

  \If{$i=1$} \Comment{Bottom layer}
    \label{line:bottom_layer}
    \State $T_2^* \gets \min(U,P)$
    \State \Return
  \EndIf

  \If{$U<P$} \Comment{Within total power $P$}
    \State Compute $W$ from $U,w,\beta,u,\alpha$
    \State \Call{alloc}{$i-1,W,\alpha,p_{i-1},\gamma_{i-1}$} \Comment{Unconstrained}
      \label{line:alloc_uncon}
    \If{$T_i^* \geq U$}
      \State $T_{i+1}^* \gets U$ \Comment{$U$ is feasible}
      \State \Return
    \EndIf
  \EndIf

  \State \Call{alloc}{$i-1,w,\beta,p_{i-1}+u,\gamma_{i-1}$} \Comment{Constrained}
    \label{line:alloc_con}
  \State $T_{i+1}^* \gets T_i^*$

\EndProcedure
\end{algorithmic}
\end{algorithm}

\bibliographystyle{IEEEtran}
\bibliography{IEEEabrv,wrlscomm}

\begin{biographynophoto}
{Chris~T.~K.~Ng} received his B.A.Sc. in Engineering Science from the University of Toronto. He received his M.S. and Ph.D. in Electrical Engineering from Stanford University.
Since 2009, he has been a Member of Technical Staff at Bell Laboratories, Alcatel-Lucent, in Holmdel, New Jersey.
From 2007 to 2008, he was a postdoctoral researcher in the Department of Electrical Engineering and Computer Science at the Massachusetts Institute of Technology.
He was a recipient of the 2007 IEEE International Symposium on Information Theory Best Student Paper Award, and a recipient of a Croucher Foundation Fellowship in 2007.
His research interests include cooperative communications, joint source-channel coding, cross-layer wireless network design, optimization, and network information theory.
\end{biographynophoto}

\begin{biographynophoto}
{Deniz G\"{u}nd\"{u}z} received the B.S. degree in electrical and electronics engineering from the Middle East Technical University in 2002, and the M.S. and Ph.D. degrees in electrical engineering from Polytechnic Institute of New York University (formerly Polytechnic University), Brooklyn, NY in 2004 and 2007, respectively. He is currently a consulting Assistant Professor at the Department of Electrical Engineering, Stanford University and a postdoctoral Research Associate at the Department of Electrical Engineering, Princeton University. In 2004, he was a summer researcher in the laboratory of information theory (LTHI) at EPFL in Lausanne, Switzerland.

Dr.\ G\"{u}nd\"{u}z is the recipient of the 2008 Alexander Hessel Award of Polytechnic University given to the best PhD Dissertation, and a coauthor of the paper that received the Best Student Paper Award at the 2007 IEEE International Symposium on Information Theory. His research interests lie in the areas of communication theory and information theory with special emphasis on joint source-channel coding, cooperative communications, network security and cross-layer design.
\end{biographynophoto}

\begin{biographynophoto}
{Andrea Goldsmith} is a professor of Electrical Engineering at Stanford University, and was previously an assistant professor of Electrical Engineering at Caltech.
She is also founder and CTO of Quantenna Communications, Inc., and has previously held industry positions at Maxim Technologies, Memorylink Corporation, and AT\&T Bell Laboratories.
Her research includes work on wireless information and communication theory, MIMO systems and multihop networks, sensor networks, cross-layer wireless system design, and wireless communications for distributed control.
She is author of the book ``Wireless Communications'' and co-author of the book ``MIMO Wireless Communications,'' both published by Cambridge University Press.
She received the B.S., M.S. and Ph.D. degrees in Electrical Engineering from U.C. Berkeley.

Dr.\ Goldsmith is a Fellow of the IEEE and of Stanford.
She has received several awards for her research, including the National Academy of Engineering Gilbreth Lectureship, the Alfred P. Sloan Fellowship, the Stanford Terman Fellowship, the National Science Foundation CAREER Development Award, and the Office of Naval Research Young Investigator Award.
In addition, she was a co-recipient of the 2005 IEEE Communications Society and Information Theory Society joint paper award.
Dr.\ Goldsmith currently serves as associate editor for the IEEE Transactions on Information Theory and as editor for the Journal on Foundations and Trends in Communications and Information Theory and in Networks.
She previously served as an editor for the IEEE Transactions on Communications and for the IEEE Wireless Communications Magazine, as well as guest editor for several IEEE journal and magazine special issues.
Dr.\ Goldsmith participates actively in committees and conference organization for the IEEE Information Theory and Communications Societies and has served on the Board of Governors for both societies.
She is a Distinguished Lecturer for the IEEE Communications Society, the president of the IEEE Information Theory Society, and was the technical program co-chair for the 2007 IEEE International Symposium on Information Theory.
She also founded the student committee of the IEEE Information Theory society, is an inaugural recipient of Stanford's postdoc mentoring award, and will serve as Stanford's faculty senate chair in 2009/2010.
\end{biographynophoto}

\begin{biographynophoto}
{Elza Erkip} received the B.S. degree in electrical and electronic engineering from the Middle East Technical University, Ankara, Turkey and the M.S and Ph.D. degrees in electrical engineering from Stanford University, Stanford, CA\@.
Currently, she is an associate professor of electrical and computer engineering at the Polytechnic Institute of New York University.
In the past, she has held positions at Rice University and at Princeton University.
Her research interests are in information theory, communication theory and wireless communications.

Dr.\ Erkip received the NSF CAREER Award in 2001, the IEEE Communications Society Rice Paper Prize in 2004, and the ICC Communication Theory Symposium Best Paper Award in 2007. She co-authored a paper that received the ISIT Student Paper Award in 2007.
Currently, she is an Associate Editor of IEEE Transactions on Information Theory, an Associate Editor  of IEEE Transactions on Communications, the Co-Chair of the GLOBECOM 2009 Communication Theory Symposium and the Publications Chair of ITW 2009, Taormina.
She was a Publications Editor of IEEE Transactions on Information Theory during 2006--2009, a Guest Editor of of IEEE Signal Processing Magazine in 2007, the MIMO Communications and Signal Processing
Technical Area Chair of the Asilomar Conference on Signals, Systems, and Computers in 2007, and the Technical Program Co-Chair of Communication Theory Workshop in 2006.
\end{biographynophoto}

\end{document}